\documentclass[12pt]{iopart}
\usepackage{iopams} 
\usepackage{graphicx}

\usepackage[usenames,dvipsnames,svgnames]{xcolor}

\expandafter\let\csname equation*\endcsname\relax
\expandafter\let\csname endequation*\endcsname\relax
\usepackage{amsmath}

\usepackage{multirow}
\usepackage{booktabs}
\usepackage{siunitx}

\newcommand{\defeq}{\stackrel{\text{\tiny def}}{=}}
\usepackage{comment}

\begin{document}

\title[A study of OPMs for use in MEG without shielding]{A study of scalar optically-pumped magnetometers for use in magnetoencephalography without shielding}

\author{Richard J. Clancy$^{1}$, Vladislav Gerginov$^{2}$, Orang Alem$^{2,3}$, Stephen Becker$^{1}$, and Svenja Knappe$^{2,3}$}
\address{$^1$Department of Applied Mathematics, University of Colorado, Boulder, CO 80309}
\address{$^2$Department of Mechanical Engineering, University of Colorado, Boulder, CO 80309}
\address{$^3$Fieldline Inc., Boulder, CO 80301}
\ead{richard.clancy@colorado.edu}
\vspace{10pt}

\begin{abstract}
    Scalar optically-pumped magnetometers (OPMs) are being developed in small packages with high sensitivities. The high common-mode rejection ratio of these sensors allows for detection of very small signals in the presence of large background fields making them ideally suited for brain imaging applications in unshielded environments. Despite a flurry of activity around the topic, questions remain concerning how well a dipolar source can be localized under such conditions, especially when using few sensors. In this paper, we investigate the source localization capabilities using an array of scalar OPMs in the presence of a large background field while varying dipole strength, sensor count, and forward model accuracy. We also consider localization performance as the orientation angle of the background field changes. Our results are validated experimentally through accurate localization using a phantom virtual array mimicking a current dipole in a conducting sphere in a large background field. Our results are intended to give researchers a general sense of the capabilities and limitations of scalar OPMs for magnetoencephalography systems.
\end{abstract}

\maketitle

\section{Introduction}
    
    Magnetoencephalography (MEG) is a non-invasive method to image brain function with high spatial and temporal resolution \cite{supek2018MEG}. Electrical currents in the brain give rise to magnetic fields that are detectable on the exterior of the head. Exploring these MEG recordings can provide insight into the functionality and disorders of the brain.
    
    Since the early 1990s, superconducting quantum interference devices (SQUIDs) have been stalwarts in the world of MEG \cite{cohen1975MEG}. While the technology is mature and well understood, several practical considerations are incentivizing alternate technologies. To start, SQUIDs must be cooled requiring a bath of liquid helium which is expensive and subject to commodity shortages. Also, the Dewar wall thickness sets a limit on the proximity of sensors to the scalp. Since the magnetic fields induced by electrical currents in the brain are remarkably weak (hundreds of femto-Tesla at the surface of the scalp), any additional separation between the source and the sensor adversely impacts the signal-to-noise ratio (SNR) and the ability to localize brain activity. Furthermore, to fit all users, the rigid Dewars are sized to fit large human heads, severely limiting their usefulness on young children \cite{baillet2017magnetoencephalography}.  
    
    Optically-pumped magnetometers (OPMs) are atomic sensors that, under certain conditions, are capable of matching sensitivities observed with SQUIDs \cite{Dang2010scalarsens}. They operate without cryogenics, eliminating the need for bulky Dewars, and can be placed within millimeters of the scalp. OPMs open the door for wearable MEG systems that conform to a subject's head and allow for free and natural movement during scanning \cite{boto2018moving}. To rival the sensitivities of SQUIDS, OPMs must operate in the spin-exchange relaxation-free (SERF) regime \cite{kominis2003subfemtotesla, allred2002high}. However, the narrow linewidths of SERF OPMs \cite{allred2002high, happertang1977} limit the operating dynamic range to a few nano-Tesla or near zero-field background. This means magnetically shielded rooms (MSRs), and often additional large coils, are necessary to cancel any ambient fields and reduce interference that can severely limit usefulness in hospital environments.  
    
    An alternative to zero-field OPMs are total-field or scalar optically-pumped magnetometers, which are being developed in microfabricated packages \cite{schwindt2006CSAM} or with high sensitivities \cite{sheng2013subfemtotesla}. While they were first developed in the 1950s \cite{Kastler1950} and used for biomagnetic measurements since the 1970s \cite{Livanov1977}, only recently have they reached noise floors sufficient for MEG applications and been demonstrated in small packaged sensors \cite{limes2020sensitive}.
    
    For total field magnetometers, the sensor orientation does not affect the measured field value but only the achieved sensitivity. For that reason, total field magnetometers realize higher common-mode rejection ratios than their vector counterparts \cite{limes2020total}. 
    This allows for the detection of very small signals in the presence of large background fields, making them well suited for use in unshielded environments \cite{zhang2020portable, perry2020all}. Since scalar OPMs are less sensitive to their orientation, they are commonly used on moving or vibrating platforms for unexploded ordinance detection (UXO), geophysical surveying, and exploration \cite{du2017detection, david2004rival, nabighian2005historical}. Work has begun on detecting brain signals in unshielded environments \cite{limes2020sensitive, zhang2020recording}. Recently, an exciting proof-of-concept portable MEG system was constructed for use in Earth's field with scalar magnetometers~\cite{limes2020portable, limes2020total}, exhibiting their suitability for biomagnetic applications. 
    
   In this paper, we provide numerical characterizations of localization accuracy based on the variability in sensor count, sensitivity, and the fidelity of the forward model using a single dipolar source in a conducting sphere and large ambient field. We also investigate the influence of background (or bias) field angle on the localization accuracy. We illustrate the validity of our method by localizing a current dipole using a dry phantom and a scalar gradiometer array.

    \section{Models and algorithms}
   Following the general MEG method for dipole localization in a conducting sphere \cite{sarvas1987basic}, the sources are modeled as \textit{current dipoles} specified by their position, orientation, and magnitude of directional current flow. Current dipoles give rise to magnetic field patterns but are distinct from those of \textit{magnetic dipoles} which can be thought of as small bar magnets. The dipolar source is constrained to a depth consistent with the cerebral cortex and represents a neural bundle firing in unison. The terms ambient, background, and bias field are used synonymously.
    
   In this paper, we focused on static analysis but similar methods can be used for time series measurements through averaging, peak finding, Fourier transforms, and other processing tools. The aim of localization is to determine the dipole location and moment that best conforms to the measured data. In what follows, we describe a forward model, explain how it relates to the optimization problem we use for localization, and then outline an algorithm to find sources of neuronal activity. 

\subsection{Forward model} 
    A forward model characterizes the magnetic field of a known source throughout space. We consider the conducting sphere model of radius 9.1\, cm provided in \cite{sarvas1987basic} and elaborated on in \cite{mosher1999eeg}. 
    \begin{figure}
        \centering
        \includegraphics[width=.5 \textwidth]{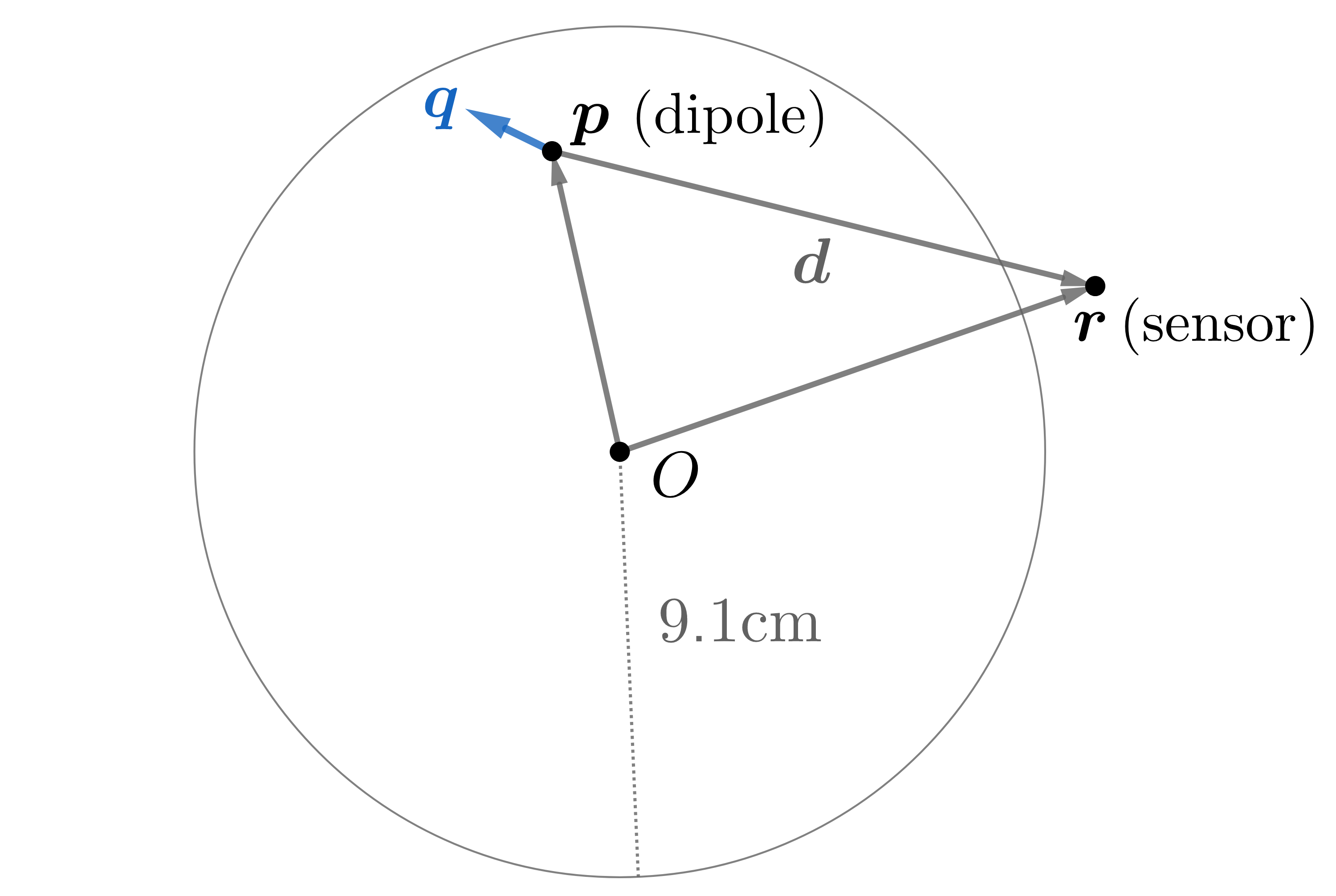}
        \caption{A current dipole at point $\bi p$ with moment $\bi q$ will generate a field at sensor location $\bi r$ according to Eq. \eqref{eq:bfield_with_kernel} and \eqref{eq:conducting_sphere_lead_field}}
        \label{fig:diagram}
    \end{figure}
    The authors show that the magnetic field at point $\bi r$ due to a dipole located at point $\bi p$ can be written as the product of a \textit{solution kernel}, $\bi L (\bi r,\bi p)$, and the dipole moment, $\bi q$. That is, for a dipole $(\bi p, \bi q)$, the magnetic field at point $\bi r$ is
    \begin{equation} \label{eq:bfield_with_kernel}
        \bi B(\bi r, \bi p, \bi q) = \bi L(\bi r, \bi p) \bi q.
    \end{equation}
    The solution kernel, $\bi L$, is given by
    \begin{equation} \label{eq:conducting_sphere_lead_field}
        \bi L( \bi r, \bi p) = \frac{\mu_0}{4 \pi} \left[ \frac{ \nabla_{\bi p} \Phi \bi r^T - \Phi \bi I}{\Phi^2} \bi C_{\bi p} \right] \qquad \text{with} \qquad
        \bi C_{\bi p} = \begin{pmatrix} 0 & -p_z & p_y \\ p_z & 0 & -p_x \\ -p_y & p_x & 0  \end{pmatrix},
    \end{equation}
    where $p_x, \ p_y,$ and $p_z$ are the components of $\bi p$, and $\bi I$ is the identity matrix. Note that $\Phi \in \mathbb{R}$, \ \  $\bi p, \, \bi q , \, \bi r, \, \bi B, \, \nabla_{\bi p}\Phi \in \mathbb{R}^3$, \ \  and $\bi C_{\bi p}, \, \bi I, \, \bi L(\bi r, \, \bi p) \in \mathbb{R}^{3 \times 3}$. Defining relative position as $\bi d = \bi r - \bi p$, the scalar function $\Phi$ and its gradient with respect to $\bi p$ are given by 
   
    \begin{align} \label{eq:nasty_f}
        \Phi(\bi r, \bi p) & = \|\bi d\| \left( \|\bi d\| \, \|\bi r\| + \|\bi r\|^2 - \bi p^T \bi r \right) \\ 
        \nabla_{\bi p} \Phi(\bi r, \bi p) &= \left( \frac{\|\bi d\|^2}{\|\bi r\|} + \frac{\bi d^T \bi r}{\|\bi d\|} + 2 \|\bi d\| + 2 \|\bi r\| \right) \bi r -  \left( \|\bi d\| + 2 \|\bi r\| + \frac{\bi d^T \bi r }{\|\bi d\|} \right) \bi p. \nonumber
    \end{align}
    In the absence of clutter, the measured signal, $y_i$, of a scalar sensor at location $\bi r_i$ will be the norm of the superposition of the ambient field, $\bi a$, and the dipolar field, $\bi B$, plus noise, $\eta_i$. That is, $y_i = \| \bi B(\bi r_i, \bi p, \bi q) + \bi a\| + \eta_i$. We consider cases where the ambient field is sufficiently large such that $y_i$ is always positive. Since the background fields we consider are many orders of magnitude larger than the biomagnetic signals we aim to measure, i.e., micro vs. femto-Telsa, we can treat scalar sensor readings as a superposition of the dipolar and ambient field, $ \bi B(\bi r_i, \bi p, \bi q) + \bi a$, projected onto the ambient field's direction. That is, letting $\hat{\bi a}$ be the ambient field's unit vector, writing $\bi b_i = \bi B(\bi r_i, \bi p, \bi q)$ and assuming $\|\bi b_i\| \ll \|\bi a\|$, the noiseless scalar magnetic field at point $\bi r_i$ is
    \begin{align} \label{eq:taylor_approx}
        \| \bi a + \bi b_i\| &= \sqrt{ \|\bi a\|^2 + 2 {\bi a}^T\bi b_i + \|\bi b_i \|^2 } \nonumber \\
        &\approx \sqrt{ \|\bi a \|^2 + 2 {\bi a}^T\bi b_i }  \nonumber \\
        &= \|\bi a\| \sqrt{ 1+2 \hat{\bi a}^T\bi b_i/\|\bi a\| } \nonumber \\
        &\approx \|\bi a\|\left( 1+\hat{\bi a}^T\bi b_i/\|\bi a\| \right) \quad\text{(Taylor series)} \nonumber \\
        &= \|\bi a\|+\hat{\bi a}^T \bi B(\bi r_i, \bi p, \bi q) \defeq f_i. \
    \end{align}
    Thus for an MEG array with $M$ sensors at points $\bi r_1, \bi r_2, ..., \bi r_M$, the forward model is 
    \begin{equation} \label{eq:forward_model}
        \bi f = \begin{bmatrix}  \hat{\bi a}^T \bi L(\bi r_1, \bi p) \\  \hat{\bi a}^T \bi L(\bi r_2, \bi p)  \\ \vdots  \\  \hat{\bi a}^T \bi L(\bi r_M, \bi p)  \end{bmatrix} \bi q + \|\bi a\| \boldsymbol{1}
    \end{equation}
    where $\boldsymbol{1}$ is the vector of ones. 
    This approximation is an affine function of $\bi q$ paving the way for a simplified localization algorithm. Upon testing, we found that the difference between our approximation and the precise forward model, i.e., $\big| f_i - \|\bi a + \bi b_i \| \big|$, was below the noise floor used and therefore negligible.

%=================================================================
\subsection{Gradiometry}
    Since we are interested in detecting weak brain signals dominated by large background fields, filtering is paramount. Outside of laboratory settings, we expect the ambient field to vary over space and time for many reasons including solar winds, passing vehicles, proximity to ferrous materials, etc. The presence of clutter necessitates gradiometry. 
    
    For simulations and experiment, we used two scalar magnetometers oriented radially as our gradiometers. A \textit{primary} sensor at $\bi r^P$ was mounted on the surface of the conducting sphere and was intended to measure brain magnetic fields. The \textit{secondary} sensor at $\bi r^S$ was employed to sense the background field but should, in principle, be far enough away to not detect biomagnetic signals. The superscripts are not exponents. The distance or baseline between primary and secondary sensors is 4\, cm in the radial direction. For this gradiometer arrangement, the forward model corresponding to measured difference data, $y^P - y^S$, is 
    \begin{align}
        f^P - f^S \ \ = & \ \  (\hat{\bi a}^T \bi L(\bi r^P, \bi p) \bi q + \| \bi a \|) - (\hat{\bi a}^T \bi L(\bi r^S, \bi p) \bi q + \| \bi a \|) \\
        = & \ \ \hat{\bi a}^T\big(\bi L(\bi r^P, \bi p) - \bi L(\bi r^S, \bi p) \big) \bi q,
    \end{align}
    which has no dependence on the magnitude of the bias field. The full gradiometer forward model can be written as 
    \begin{equation} \label{eq:gradiometer_forward_model}
        \bi f^G(\bi p, \bi q) \ \ = \ \ \begin{bmatrix}  \hat{\bi a}^T \left( \bi L(\bi r^P_1, \bi p) - \bi L(\bi r^S_1, \bi p) \right) \\  \hat{\bi a}^T \left( \bi L(\bi r^P_2, \bi p) - \bi L(\bi r^S_2, \bi p) \right) \\ \vdots \\ 
        \ \hat{\bi a}^T \left( \bi L(\bi r^P_M, \bi p) - \bi L(\bi r^S_M, \bi p) \right) \\  \end{bmatrix} \bi q \ \ = \ \ \bi A(\bi p) \, \bi q.
    \end{equation}
    The superscript denotes the gradiometer arrangement forward model. For the matrix $\bi A(\bi p)$, we treat the sensor locations as fixed henceforth. For primary and secondary sensor readings $\bi y^P$ and $\bi y^S$, respectively, the corresponding gradiometer measurement vector is given by 
    \begin{equation} \label{eq:gradiometer_measurement}
        \bi y^G = \bi y^P - \bi y^S.
    \end{equation}
    With a forward model and data at our disposal, we can now formulate the optimization problem to localize a dipole.

%===================================================================
\subsection{Optimization problem and algorithms} \label{sec:algorithms}
    Given the forward model $\bi f^G$ from \eqref{eq:gradiometer_forward_model}, the goal is to find a dipole $(\bi p, \bi q)$ that best explains observed MEG data $\bi y^G$. We formulate this mathematically with the nonlinear least squares problem 
    \begin{equation} \label{eq:optimization_problem}
        \min_{\bi p,\bi q} \ \| \bi f^G(\bi p, \bi q) - \bi y^G \|^2 =\min_{\bi p, \bi q} \ \|\bi A (\bi p) \, \bi q - \bi y^G \|^2
    \end{equation}
    which is convex in $\bi q$ but not in $\bi p$. To address non-convexity, we break the problem into two parts as outlined in \cite{ilmoniemi2019brain}. First, we find a point that is in the global optimum's basin of attraction as a warm start, then employ an iterative nonlinear solver.
    
    For the warm start, we grid the variable $\bi p$
    over the sphere's interior at $K$ locations. Spacing of 2\, cm is typically sufficient although we used 1.1\, cm here. For each grid value of $\bi p$, the optimal value of $\bi q$ is easily found by solving an ordinary least squares problem, so $\bi q$ does not need to be discretized. In particular, at a fixed grid point $\bi p = \bi p_k$ with $k \in\{1,\ldots,K\}$, the optimal solution $\bi q_k^*$ to problem \eqref{eq:optimization_problem} is 
    \begin{equation} \label{eq:ols_solution}
        \bi q_k^* = \big( \bi A(\bi p_k)^T \bi A(\bi p_k) \big)^{-1} \bi A(\bi p_k)^T \bi y^G.
    \end{equation}
    After finding location/moment least square pairs, $(\bi p_k, \bi q_k^*)$, for all $K$ discrete locations, the optimal index $o$ is given by
    \begin{equation}
        o = \underset{k\in\{1,\ldots,K\}}{\text{argmin}} \ \|\bi A (\bi p_k) \, \bi q_k^* - \bi y^G \|^2.
    \end{equation}
    Hence, we use pair $(\bi p_o, \bi q_o^*)$ as a warm start for a continuous optimization algorithm. In our simulations, we used L-BFGS \cite{liu1989limited} as the continuous optimization algorithm, which is a quasi-Newton method, although others could be used. L-BFGS iterates through the space of possible values for $(\bi p, \bi q)$ to find a fit minimizing residual error.

%===================================================================   
\section{Numerical Experiments} \label{sec:numerical_experiments}
    We ran a variety of simulations to understand limitations and requirements for scalar OPM use in MEG systems with strong ambient fields. In all simulations we use gradiometry. As such, the sensor count reflects the number of gradiometers, not the number of sensors (each gradiometer has a primary and secondary sensor). We provide multiple curves in each experiment indicating performance for different \textit{relative dipole strengths} (RDS) defined by 
    \begin{equation}
    \text{RDS}= \frac{\text{Dipole strength (nAm)}}{ \bigg(  \text{Sensor sensitivity (fT/}\sqrt{\text{Hz}}\,) \bigg) \bigg( \sqrt{\text{Bandwidth (Hz)}} \bigg)}.
    \end{equation}
    This figure of merit is chosen to easily relate the results to a variety of magnetometers and experiments with different noise, bandwidth, and dipole source strengths. A typical physiological current dipole generated by synchronous neural activity is 10 nAm \cite{hamalainen1993magnetoencephalography}. In our experiments, we used a scalar OPM sensor with a noise floor of 70 fT/$\sqrt{\text{Hz}}$ \cite{gerginov2020} and measurement bandwidth of 100 Hz. For these OPM parameters, we then varied the dipole strength for simplicity in our simulation. A 10 nAm dipole gives an RDS of 0.014 nAm/fT. See Table \ref{tab:conversion} for conversions between bandwidths and sensitivities.
    
    \begin{table}
    \centering
    \begin{tabular}{c c | c c c c}
    \toprule
        RDS  & & \multicolumn{4}{c}{\textbf{Sensor sensitivity (fT/$\sqrt{\text{Hz}}$)}} \\
        & & 0.1 & 1.0 & 10.0& 100.0 \\ 
        \hline
        \multirow{7}{1em}{\rotatebox{90}{\textbf{Bandwidth (Hz)}}} &
        1& 100.000& 10.000& 1.000& 0.100 \\
        &5& 44.721& 4.472& 0.447& 0.045 \\
        &10& 31.623& 3.162& 0.316& 0.032 \\
        &50& 14.142& 1.414& 0.141& 0.014 \\
        &100& 10.000& 1.000& 0.100& 0.010 \\
        &500& 4.472& 0.447& 0.045& 0.004 \\
        &1000& 3.162& 0.316& 0.032& 0.003
    \end{tabular}
            
    \caption{Relative dipole strength (RDS) in units of nAm/fT for a physiological neuronal current dipole of magnitude 10 nAm over different bandwidths and sensitivities.}
    \label{tab:conversion}
    \end{table}

    Each experiment consists of 10,000 dipoles randomly drawn from a specified volume. All dipoles in this paper have a random orientation that is tangential to the spherical surface. To generate scalar sensor data for a ``true'' simulated dipole $(\bi p_\text{T}, \bi q_\text{T})$, we first calculate the dipolar field from Equations \ref{eq:bfield_with_kernel}, \ref{eq:conducting_sphere_lead_field}, and \ref{eq:nasty_f} then add it to the ambient field. We take the magnetic field norm at all sensor locations (both primary and secondary) to give noiseless sensor readings. A random Gaussian vector with zero mean and standard deviation of 700 fT (70 fT/$\sqrt{\text{Hz}}$ times $\sqrt{100 \text{Hz}}$) is drawn and added to the noiseless measurement vector. Finally, we subtract the secondary from the primary sensor readings to give the gradiometer measurement vector in \eqref{eq:gradiometer_measurement}. We can then employ the algorithm outlined in Section \ref{sec:algorithms} with the generated gradiometer measurement vector as our input.
    
    For each simulation, the algorithm returns an estimated dipole, $(\Tilde{\bi p}, \Tilde{\bi q})$, which is compared with the true dipole through localization error $\|\bi p_\text{T} - \Tilde{\bi p} \|$. Although we must estimate $\bi q$, we are not concerned with it and focus on dipole location error alone. We use median and other quantile based error over 10,000 simulations to evaluate accuracy. 
    
\begin{figure}
        \centering
        \includegraphics[width=.6 \textwidth]{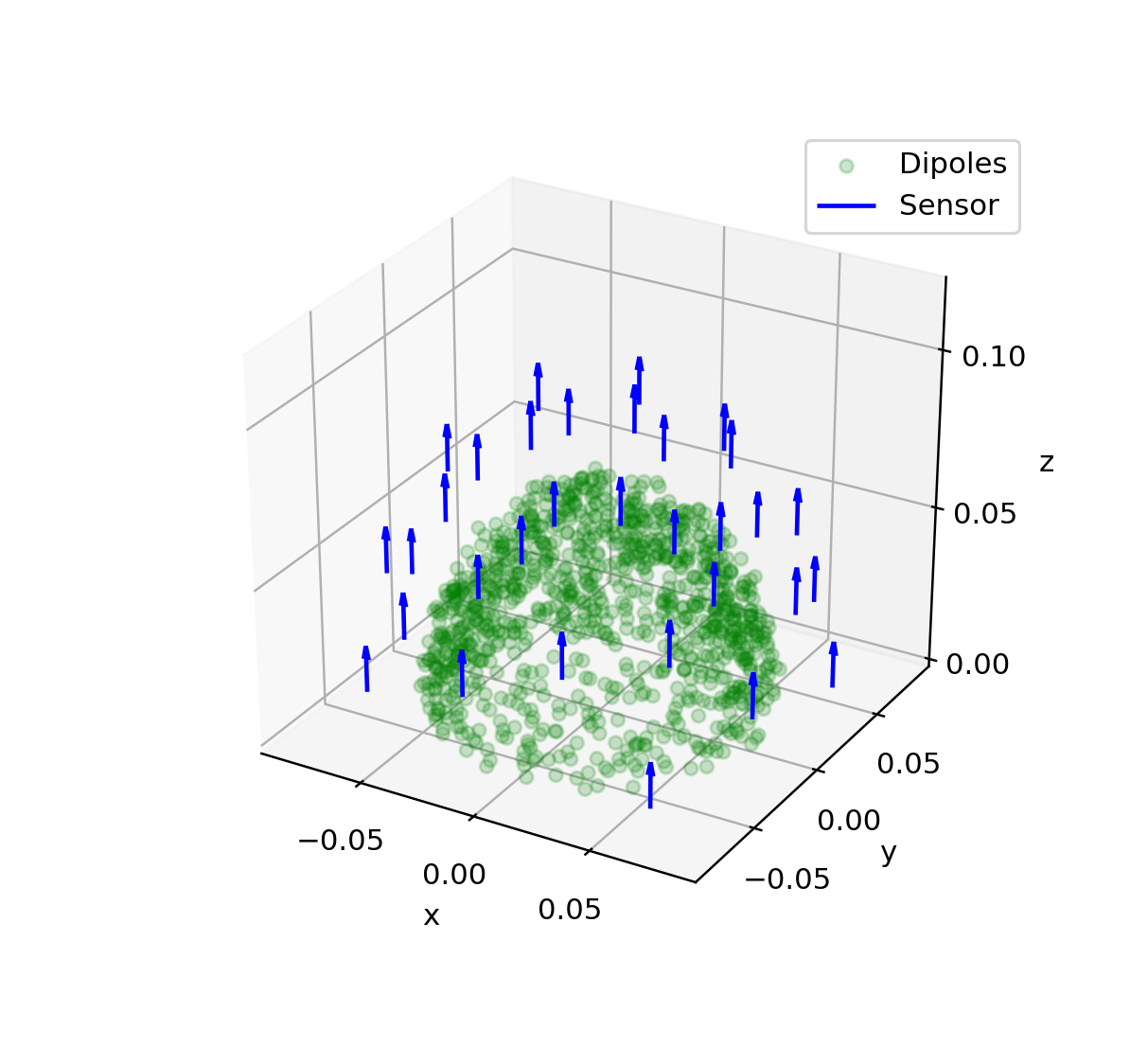}
        \caption{Typical 32 sensor arrangement using Fibonacci spiral with bias field along $z$-axis. Green dots are locations of simulated dipoles drawn randomly with surface depth of 2 -- 3.5\, cm. Scale in meters.}
        \label{fig:typical_sensor_arrangement}
        % plots and experiments can be found in gradiometer_dipole_strength folder
    \end{figure}

    Except for bias field orientation simulations, sensors were placed on the upper half of a hemisphere along a Fibonacci spiral \cite{vogel1979better, gonzalez2010measurement} which is a parametric curve giving nearly uniform coverage over a spherical surface. The lower boundary of the hemisphere is the $xy$-plane. We outline our numerical simulations in detail below. A typical sensor arrangement is shown in Figure \ref{fig:typical_sensor_arrangement}. Since we assume scalar sensors, the orientation of the OPM is not important and the blue arrows denote the direction of the large bias field assumed at the location of the sensor.

%===========================================================    
\subsection{Vector vs. scalar sensor comparison} \label{experiment:sensor_comparison}
    Usually, MEG is recorded with directional sensors, measuring a single component of the magnetic field vector. We will denote them as ``vector'' sensors for simplicity, even though the full vector field is not measured. In contrast, scalar sensors measure the total magnitude of the field independent of orientation. 
    
\begin{figure}
        \centering
        \includegraphics[width=.6 \textwidth]{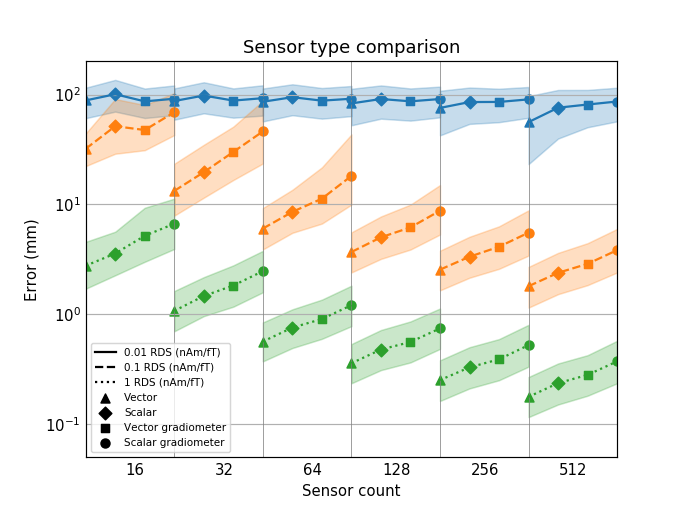}
        \caption{Each curve shows progression of error by sensor type. From left to right, sensor types are vector, scalar, vector gradiometer, and scalar gradiometer. Median localization error and middle 50th quantile error (shaded region) vs. sensor types. Each curve corresponds to a different relative dipole strength (RDS) of given sensor type and varies from a 16 to 512 sensor arrangement.}
        \label{fig:sensor_comparison}
        % plots and experiments can be found in gradiometer_dipole_strength folder
    \end{figure}

    To give the reader a better idea of how scalar and vector sensors compare, Figure \ref{fig:sensor_comparison} shows the median and middle 50th quantile error for vector and scalar sensors, as well as their gradiometry counterparts.
    We varied the number of sensors and RDS. For each simulation, the same dipole-generated fields measured by all four sensor varieties. Vector sensors were oriented normal to the conducting sphere surface, detecting radial components of the magnetic field. For both scalar sensor varieties, we assume that the orientation of the background field is provided to the forward model. Dipoles were drawn uniformly at random within the upper half of the hemisphere at a depth of 2 -- 3.5 cm to mimic activity on the cortex and with random tangential orientations.
    
    Both vector sensor arrangements perform better than their scalar versions, independent of sensor count and dipole strength. Nevertheless, it can be seen that the difference is about a factor of two and that sensor count and SNR have a much larger effect on the localization error than the type of sensor.

%===========================================================    
\subsection{Dependence on sensor count} \label{experiment:sensor_count}
    Increasing the number of sensors ostensibly improves localization accuracy. Simulations were conducted to quantify the impact of increased sensor count. Using the same setup as detailed in Section \ref{experiment:sensor_comparison}, we varied the number of scalar gradiometers from 16 to 512 in powers of 2 and recorded the localization error for each dipole simulated.  
    \begin{figure}
        \centering
        \includegraphics[width=.6\textwidth]{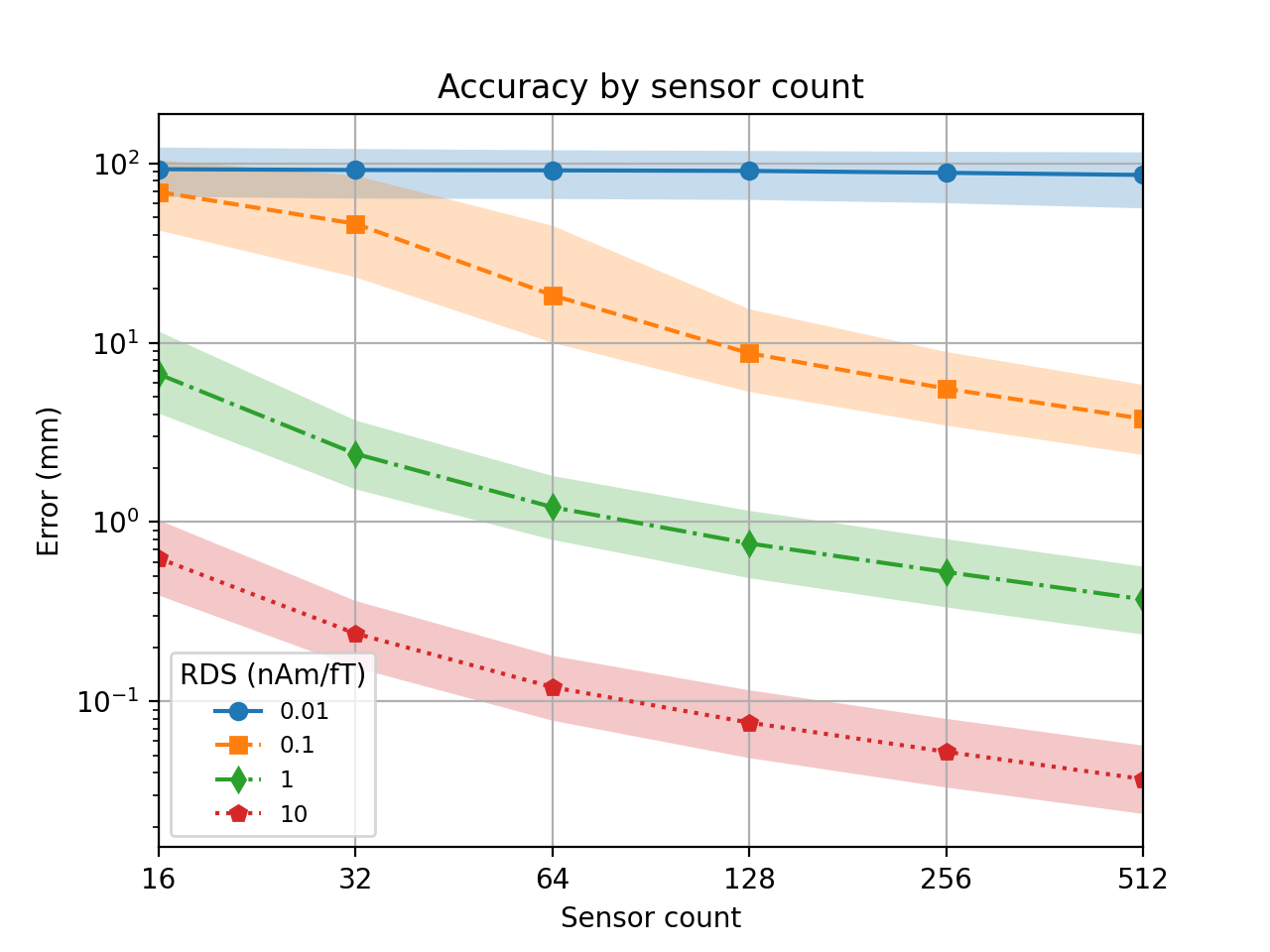}
        \caption{Median localization error (curve) and middle 50th quantile error of scalar gradiometers (shaded region) vs. sensor count for different RDS. Bias field oriented in positive $z$-direction.}
        \label{fig:sensor_count}
        % plots and experiments can be found in gradiometer_dipole_strength folder
    \end{figure}
    
    Figure \ref{fig:sensor_count} illustrates localization accuracy as a function of sensor count. As intuition suggests, more sensors improve localization accuracy. It can be observed, however, that increasing the number of gradiometers does not improve performance for the 0.01 nAm/fT curve over the sensor counts considered. It is important to achieve a baseline sensitivity before adding more sensors since they will not give the desired improvement. \emph{We estimate that localization error of 1\, cm can be reached for a 10 nAm dipole using 128 sensors with a noise floor of 10 fT/$\sqrt{\text{Hz}}$ and a bandwidth of 100 Hz. Likewise, a 1\, mm localization error is achievable with 80 sensors of similar noise floor in a 1 Hz bandwidth}. While these simulations omit many practical considerations, they give an idea of the system complexity needed to achieve reasonable localization results.

%===========================================================    
\subsection{Sensitivity to perturbations}
    Localization error depends heavily on the accurate specification of a forward model. In practice, it is challenging to determine sensor location and orientation precisely, especially since an advantage of uncooled sensors is that they can be placed in conformal geometries individual to every subject. Accordingly, we'd like to understand how robust localization is to model misspecification. To accomplish this, we fixed a presumptive sensor array for use in the forward model. We then perturbed sensor locations and bias field orientation by adding Gaussian noise to give a \textit{true} sensor array. 
    The standard deviations used varied from 0.1º to 3º for orientation and 0.1\, mm to 3\, mm for sensor location. 
    Perturbations to orientation were common across all sensors; this noise model reflects uncertainty in the direction of a subject's head in a uniform bias field rather than variations of the bias field over space. Dipoles were drawn as in Section \ref{experiment:sensor_comparison}. Finally, we generated true data by using a forward model based on the perturbed sensors and the randomly drawn dipole. For each simulation, we calculated localization error using the presumptive forward model. 

\begin{figure}
        \centering
        \includegraphics[width=.6\textwidth]{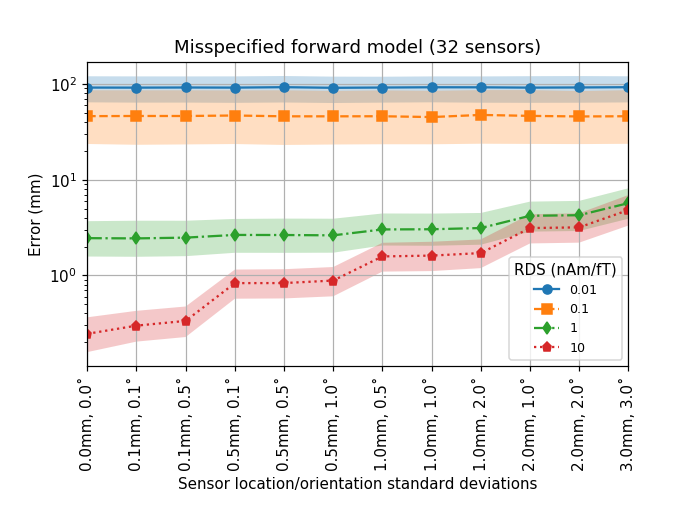}
        \caption{Median localization accuracy (curve) and middle 50th quantile error for a 32 gradiometers with varied perturbations to the sensor array used in forward model specification.}
        \label{fig:err_by_pert_32sensors}
        % plots and experiments can be found in gradiometer_perturbations folder
    \end{figure}

    Localization error as a function of perturbation level for an array with 32 gradiometers is shown in Figure \ref{fig:err_by_pert_32sensors}. These results are consistent across changes in array size. Perturbations impact the accuracy of sensitive arrays more. Small RDS values of 0.01 and 0.1 perform poorly even with correctly specified forward models, hence, there is little accuracy to lose as perturbation levels increase. 
    
    Correctly specifying sensor locations within 0.1\, mm seems challenging for a wearable scalar OPM MEG system. Commercially available optical scanners can localize sensors to 0.5\, mm, but the question of how well sensors remain in place during a recording persists, especially if the subject can move their head. Simulations show that scalar OPMs are fairly robust to perturbations in bias field orientation, but more sensitive to the uncertainty in sensor location. Sensitivity to sensor uncertainty is impacted by sensor count and dipole strength as well. In the case of 32 sensors  (Fig. \ref{fig:err_by_pert_32sensors}), increased forward model accuracy has little impact on dipoles with low RDS. It is therefore imperative to increase sensitivity to a point before adding more sensors to improve localization accuracy. 
    
%===========================================================        
\subsection{Bias field angle dependence}
    Vector sensors are frequently aligned normal to the subject's head. With scalar sensors in a bias field, sensor orientation is typically fixed by the ambient field, e.g., Earth's field, independent of sensor orientation. On the other hand, the sensor array itself can change its orientation relative to this field by rotation. How does localization accuracy depend on the relative direction of the bias field? Are certain orientations better or worse?
    
\begin{figure}
        \centering
        \includegraphics[width=.75\textwidth]{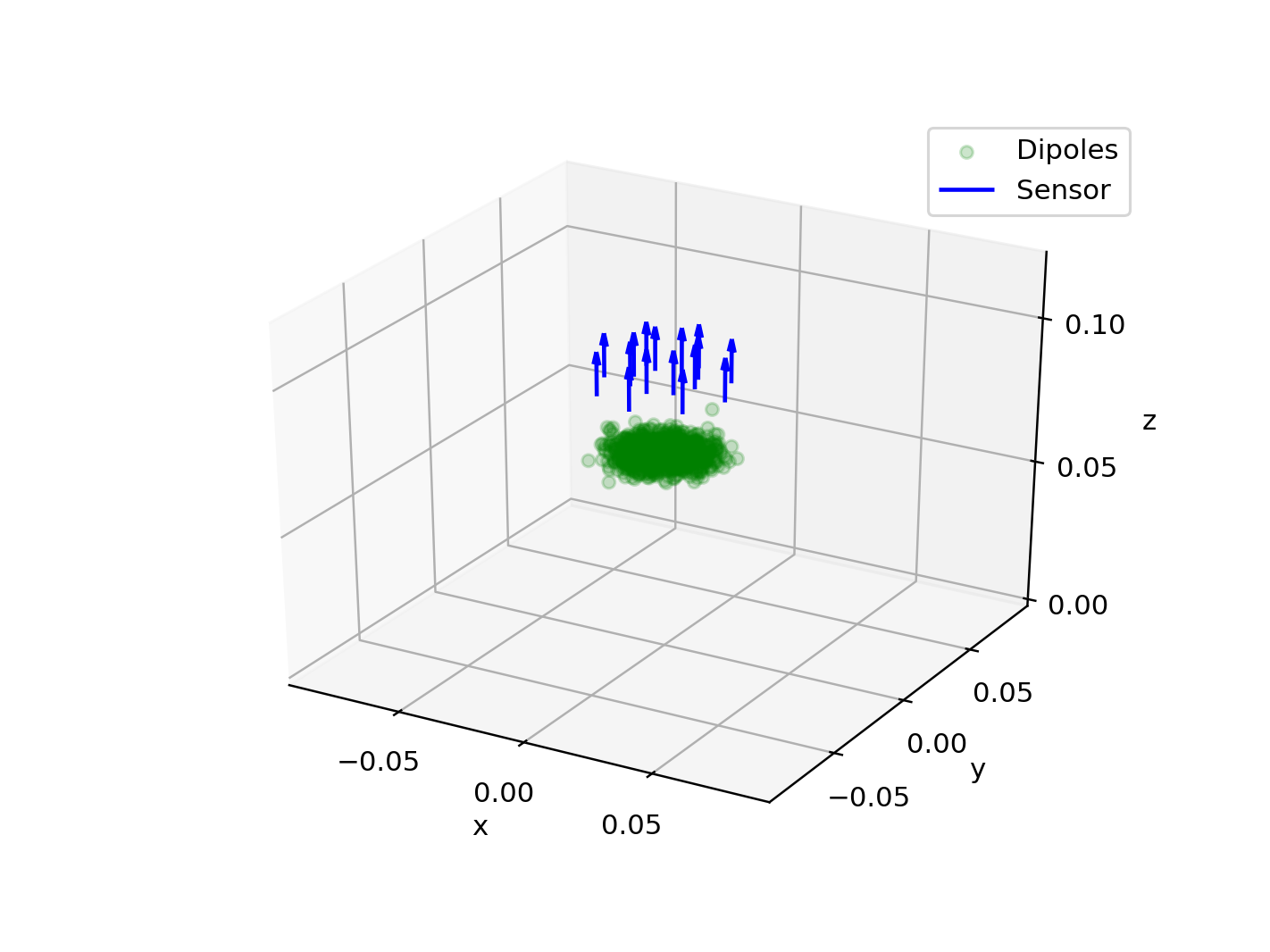}
        \caption{Typical arrangement for bias field orientation experiments. Blue arrows indicate location of sensors and effective bias field orientation at that location, green dots mark realizations of randomly drawn dipoles used to generate data. All moment oriented randomly, but tangential to spherical surface. Scale in meters.}
        \label{fig:bias_field_arrangement.png}
        % plots and experiments can be found in bias_field_experiment folder
    \end{figure}

\begin{figure}
        \centering
        \includegraphics[width=.6\textwidth]{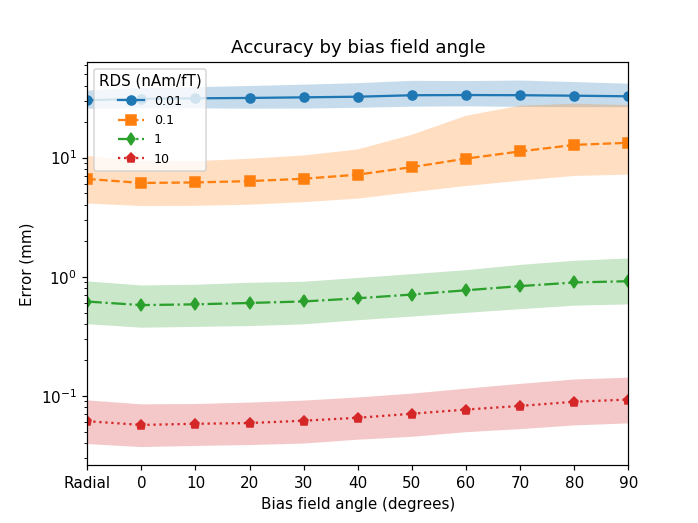}
        \caption{Median (curve) and middle 50th quantile (shaded region) localization error as a function of bias or ambient field for different RDS. Radial arrangement measures radial component of field at each sensor. Accuracy deteriorates slightly as bias field angle increases.}
        \label{fig:bias_field.png}
        % plots and experiments can be found in bias_field_experiment folder
    \end{figure}

    We assume here that the sensors are not sensitive enough to detect a signal from brain regions on the opposite side of the head. We therefore arrange a dense sensor array over a localized area of the head. Sixteen gradiometers are arranged on the hemisphere's surface in two concentric circles about the $z$-axis. The circles have azimuthal angles of 7.5º and 15º with gradiometers spaced every 45º in the polar direction. Dipoles were drawn from a region roughly below the sensor array at a mean depth of 2.5\, cm. A typical sensor arrangement with a subset of simulated dipole locations is shown in Figure \ref{fig:bias_field_arrangement.png}.
    For a simulated dipole, we attempted localization for different bias field orientations varied from 0º (along the $z$-axis) to 90º (along the $x$-axis) in 10º increments. For comparison to typical radial vector sensor arrangements, we also simulated radially mounted sensors in the same locations. All non-radial bias field orientations varied in the $xz$-plane without a $y$ component. Figure \ref{fig:bias_field.png} shows accuracy as a function of the bias field.
    
    Based on our simulations, performance deteriorates slightly as the bias field differs from the axis of symmetry around which dipoles are drawn, at least when dipoles are tightly bunched. It is also worth noting that radial sensors perform slightly worse than 0º orientation. We believe this is due in part to the fact that dipoles close to the $z$-axis generate smaller projections on average along the radial direction and hence have lower RDS. We conclude that it will be possible to image with any bias field orientation and that the localization accuracy degrades by less than a factor of two between radial and tangential bias field.

%===========================================================        
\section{Phantom experiment} \label{sec:phantom}
   
\begin{figure}
        \centering
        \includegraphics[width=.6\textwidth]{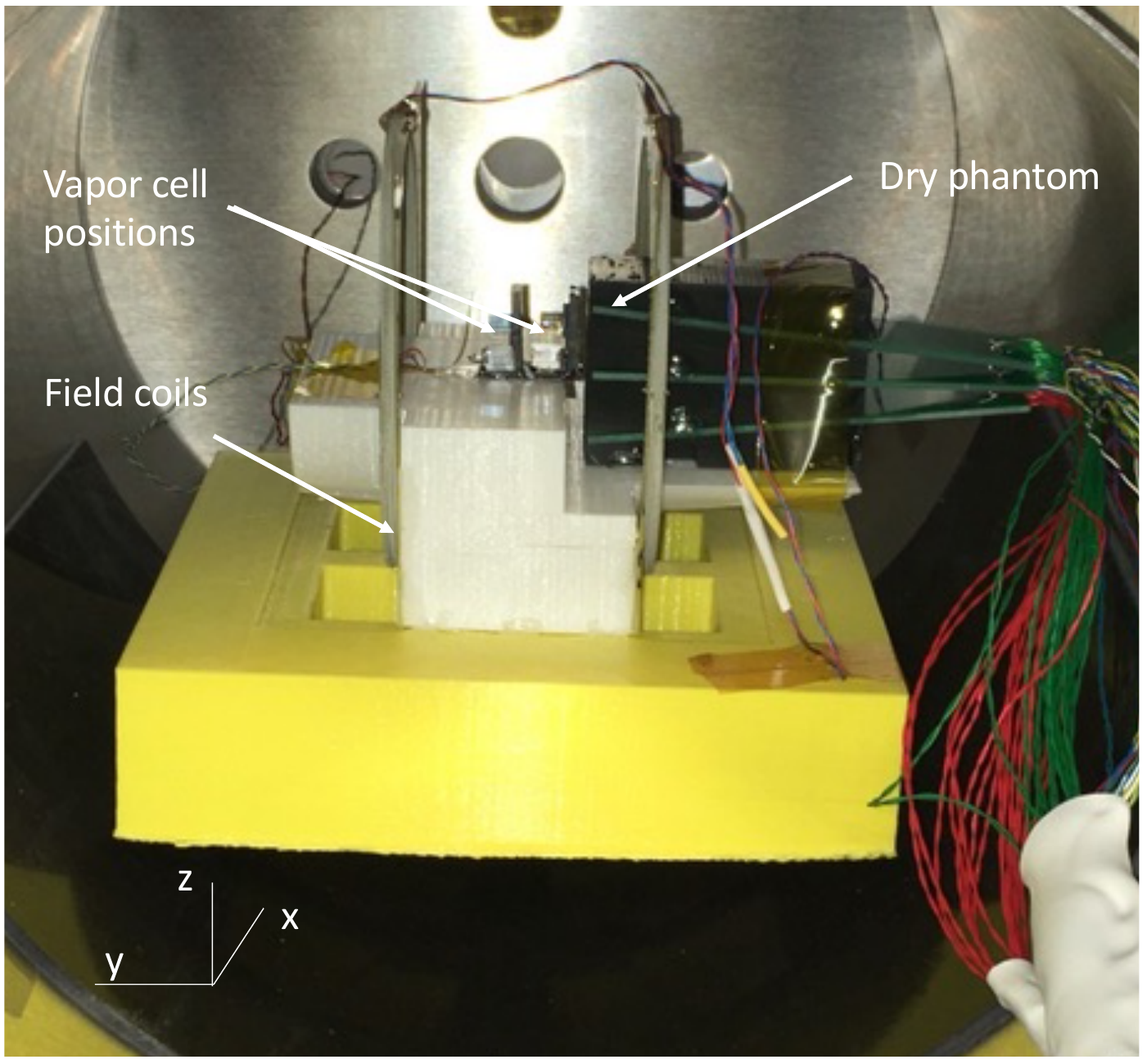}
        \caption{Phantom experiment setup.}
        \label{fig:phantomexp}
    \end{figure}

    To validate our method, we constructed a dry phantom to mimic a current dipole in a conducting sphere \cite{Ilmoniemi, Oyama, Uehara}. We constructed a virtual sensor array based on a single scalar gradiometer using a scalar magnetometer concept described in detail in Ref.~\cite{gerginov2020}. The gradiometer had a fixed base distance and was used to record many dipolar sources spread over a volume consecutively. The gradiometer setup is shown schematically in Figure~\ref{fig:phantomexp}. Two glass-blown vapor cells with $3\times4.5\times5$\, mm$^3$ internal volume were filled with isotopically-pure Rubidium $^{87}$Rb and 600\, Torr Nitrogen N$_2$ acting as a buffer gas. The vapor cells were heated with miniature non-magnetic electric heaters to approximately 373\,K. The cells were positioned 15\,mm apart inside a three-layer magnetic shield. Circularly-polarized laser beams from the same laser entered the vapor cells through the $3\times5$\,mm$^2$ cell entrance ports ($y$-axis). The laser frequency was on resonance with the $D_1$ $^{87}$Rb atomic transition that was broadened by Rb collisions with the $N_2$ buffer gas. The laser peak power was $\sim$50~mW per beam. Linearly-polarized probe beams from the same laser entered each vapor cell through the $3\times4.5$\,mm$^2$ cell entrance ports ($x$-axis). The orientation of the pump and probe beams inside the gradiometer vapor cells was different than the scalar magnetometer of Ref.~\cite{gerginov2020}, for which the pump and probe beams were co-propagating. The probe laser was detuned by $\sim$40~GHz from the $D_2$ $^{87}$Rb atomic transition. The transmitted probe light of $\sim$2~mW per beam was detected with two balanced polarimeters. A static magnetic field of 10\,$\mu$T was generated using Helmholtz coils of 50\,mm radius in the $y$-direction orthogonal to both probe and pump beams. The magnetometers were operated by AM pump light modulation (100\% depth, 3\% duty cycle at a frequency corresponding to half the Larmor frequency). The polarimeter signals were demodulated with a phase-sensitive detector \cite{BellBloom1977}. Each magnetometer had a white-noise floor of 70 fT/$\sqrt{\text{Hz}}$ in a 500~Hz bandwidth. The magnetometer arms were calibrated separately by stepping the pump light modulation frequency by 10 Hz and measuring the voltage change at the phase-sensitive detector output. The magnetometer outputs were recorded individually, calibrated, and subtracted in software.
    
    The dipolar sources were approximated by currents flowing along the boundary of arcs with small subtended angle \cite{Ilmoniemi} fabricated on a Electroless nickel immersion gold (ENIG) printed circuit board (PCB). The dipoles each had a length of 1\,mm. They were placed on a circle of radius 70\,mm and connected with thin twisted wires at the center point of the circle. Neighboring dipoles on each PCB were separated by 1.42\,mm center to center. They were activated independently with a current source. Three identical PCBs were stacked together to place the dipoles on a sphere with radius 100\,mm, each PCB separated by 12.8\,mm at the location of the dipoles. The phantom was placed such that the central dipole was 11.7\, mm from the closer vapor cell of the scalar gradiometer.

    The virtual array exploits the fact that magnetic fields induced by a current dipole at a point in space depends on relative and not absolute position. As such, we can relocate each dipole to a single point in space and vary sensor locations to mimic coverage over a subjects head (assuming the same dipole strength and orientation at both). For example, with a sensor at point $\bi s$ and dipoles at locations $\bi r_1$ and $\bi r_2$, displacements are given by $\bi d_1= \bi s - \bi r_1$ and $\bi d_2= \bi s - \bi r_2$. If we treat $\bi r_1$ as the true dipole location then we can write $\bi d_2=\bi s - \bi r_2 + (\bi r_1 - \bi r_1) = (\bi s + \bi r_1 - \bi r_2) - \bi r_1 = \Tilde{\bi s} - \bi r_1$. Geometrically, the two are equivalent but in the latter, we treat the dipole as fixed rather than the sensor. Care must be taken to adjust for the orientation of each dipole relative to the bias field. 
    
    For the experiment, we passed a 150 Hz alternating current through each triangular wire loop to mimic a dipole consecutively. A static bias field of  10 $\mu$T was applied along the $y$-axis of the gradiometer. The resultant field from each dipolar source was recorded for 100 seconds. This process was repeated at each dipole location in the array. By using the same signal at each location, we created a virtual array discussed prior. 
    
    Although the 150 Hz periodic signal was common to all gradiometers in the virtual array, the phase differed for each virtual array sensor. To localize, the signal of interest had to be aligned in time (peak signal strengths must occur simultaneously). To accomplish this, we took the Fourier Transform of the time series data at each sensor to recover amplitude and phase. That is, for time series $x(t)$, we took $\hat{ x}(\omega) = \mathcal{F}\{ x(t) \}$. We found the phase, $\phi$, of the $150$ Hz signal then shifted \textit{all} Fourier coefficients by $-\phi$, i.e., $\hat{x}_S(t) = \hat{x}(t) e^{-i \phi}$. Once equipped with phase-aligned data, we reconstructed our phase-shifted time series with $x_S(t) = \mathcal{F}^{-1} \{ \hat{x}_S(\omega) \}$. Performing this same procedure for all sensors in the virtual array, we aligned our time series data for localization. 
    
    We picked a grid of 9 out of the 3 x 17 dipoles in our array. To mimic a realistic geometry, the nine dipoles were spaced by roughly 12.5\, mm in $x$- and $z$-directions. We used several different band-pass filters centered at 150 Hz to process the phase-aligned time series data. Each filter reflects the precision of knowledge about the source's signature with narrow bandwidths corresponding to higher accuracy. We chunked the data set into single period subsets, then selected the time entry from each subset with the highest power across sensors. We used these high-power data points which typically correspond to peaks/troughs to localize with. 
    
\begin{figure}
        \centering
         \includegraphics[width=.8\textwidth]{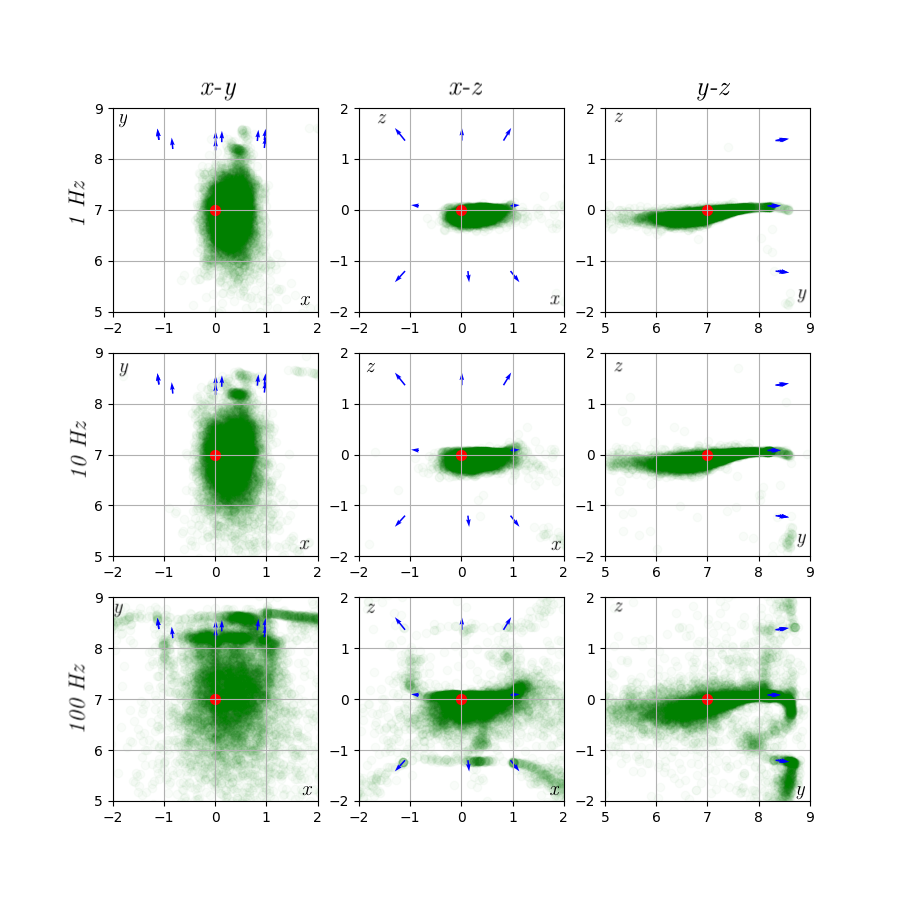}
        \caption{Elevation views of localization results for phantom array. Heavy red dots indicate true dipole location, light green dots show dipole location estimates for 10k field measurements. Blue arrows show sensor locations and orientations. Scale in centimeters.}
        \label{fig:phantom_results}
        % plots and experiments can be found in new_phantom_gradiometer folder
    \end{figure}

    For a dipole located at $(0,\ 7, \ 0)$ cm oriented in the negative $y$-direction with RDS of 2.98, 0.951, and 0.298, the median errors were 5.1\, mm, 5.5\, mm, and 8.5\, mm respectively. This agrees with simulation that estimates median error of approximately 7\, mm inaccuracy for a similar 9 sensor arrangement. Figure \ref{fig:phantom_results} shows elevation views of estimated dipole locations for the phantom virtual array data. Although results are consistent for 1 Hz and 10 Hz bandwidths, the accuracy deteriorates for 100 Hz bandwidth.

 \section{Conclusion}   

 In this study, we considered single-source dipole 
 localization with scalar OPM arrays in large uniform ambient fields. We investigated how localization depends on dipole strength/sensor noise, sensor count, bias field orientation, and uncertainty in forward models. We provided results of numerical simulation as general guidelines for future work on scalar OPM MEG arrays in ambient magnetic fields. Numerical results were validated experimentally by localizing a dipolar source on a phantom using a virtual scalar gradiometer array. 
 
Given current sensitivities of our scalar OPMs (around 70 fT/$\sqrt{\text{Hz}}$), localization of single neuronal dipoles at the 10 nAm level with a 100 Hz bandwidth under optimal conditions is unlikely unless the signal is averaged at least 100 times. With this level of averaging and an array of 128 sensors, localization accuracy around 1\, cm is predicted. Increasing sensitivity by a factor of 10 to 7 fT/$\sqrt{\text{Hz}}$ would allow localization accuracy of 1 mm with a 100 sensor array and 100 averages, provided that the positions of the sensors can be determined with an accuracy of 0.5\, mm. While this does not sound impossible, it is surely a challenging task. In these simulations, we assumed an ideal case, where the background fields were uniform across the sensor array, i.e., no close noise sources, and that the common-mode rejection of the gradiometer is sufficiently high to cancel all ambient noise and the localization is limited purely by sensor noise. Future work involves improved gradiometry to account for spatial variations in the ambient field and the localization of multiple dipoles. 
While there are still many challenges to overcome, the prospect of unshielded brain imaging with scalar magnetometers is exciting and would open many applications.

\section{Acknowledgements}
We thank Jeramy Hughes for valuable discussion and the Defense Advanced Research Projects Agency (DARPA) and National Institutes of Health (NIH) for financial support. 

Funding for the DARPA MINDSS project is through the N3 program under contract N6523619C8013 and the NIH through grants R43MH118154, R01EB027004, and R01NS094604. The views, opinions, and/or findings contained in this article are those of the authors and should not be interpreted as representing the official views or policies, either expressed or implied, of the Defense Advanced Research Projects Agency.

\newpage

\section*{References}
    \bibliographystyle{ieeetr}
    \bibliography{meg}
\end{document}